\documentclass[%
 reprint,
superscriptaddress,
groupedaddress,
preprintnumbers,
nobibnotes,
 amsmath,amssymb,
prb,
citeautoscript,
]{revtex4-1}

\newcommand{\sgn}{\mathop{\mathrm{sgn}}}

\usepackage{graphicx}
\usepackage{dcolumn}
\usepackage{bm}
\usepackage[mathlines]{lineno}
\usepackage[rightcaption]{sidecap}

\newcommand{\eps}{\varepsilon}

\begin{document}


\title{Fermi-edge singularity and related interaction induced phenomena in multilevel quantum dots.}

\author{A.S. Goremykina, E.V. Sukhorukov }
\affiliation{%
D\'{e}partement de Physique Th\'{e}orique, Universit\'{e} de Gen\`{e}ve, CH-1211 Gen\`{e}ve 4, Switzerland
}%

%
%

\date{\today}

\begin{abstract}

We study the manifestation of the non-perturbative effects of interaction in sequential tunnelling between a quasi-one dimensional system of chiral quantum Hall edge channels and a multilevel quantum dot (QD). We use the formal scattering theory approach to the bosonization technique to present an alternative derivation of the Fermi-edge singularity (FES) effect and demonstrate the origin of its universality. This approach allows us to address, within the same framework, plasmon assisted sequential tunnelling to relatively large dots and investigate the  interaction induced level broadening. The results are generalised  by taking into account the dispersion in the spectrum of plasmons in the QD. We then discuss their modification in the presence of neutral modes, which can be realised either in a QD with two chiral strongly interacting edge channels or in a three dimensional QD in the Coulomb blockade regime. In the former case a universal behaviour of the tunnelling rate is discovered.

\end{abstract}

\pacs{Valid PACS appear here}
\maketitle

\section{Introduction}

Fermi-edge singularity, originally discovered\cite{ohtaka} in the X-ray absorption spectra of metals, describes a divergence in the transition rate at low energies, which has a power-law dependence. There are two contributions to it, described by Mahan\cite{mahan} and Anderson.\cite{anderson} On one hand, 
the interaction between an electron and a hole in the final states increases the rate, on the other hand, the orthogonality catastrophe leads to its suppression.
These results were confirmed\cite{nozieres} and the exact solution for the strong interaction case was provided. 
Since then this phenomenon has been extensively studied both experimentally\cite{geim, *cobden,*maire1,*maire2,*ubbelohde} and theoretically\cite{matveev, kane,*ogawa,*prokofiev, *oreg, *komnik,hawrylak,muzykantskii, abanin}  in various configurations.

\begin{figure}[h]
\includegraphics[scale=0.2]{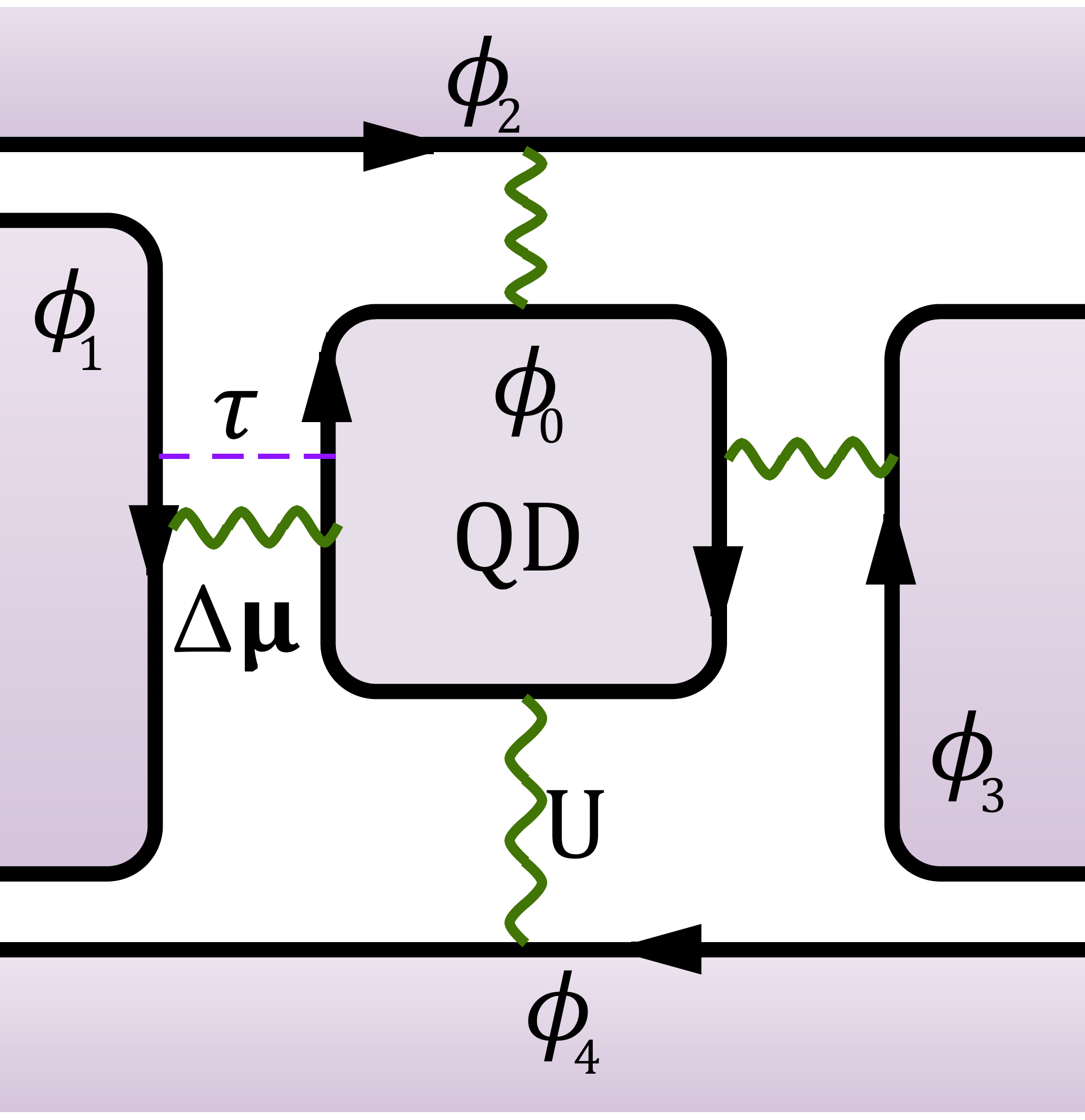} 
\caption{The scheme of the system set-up. A quantum dot, described by the bosonic field $\phi_0$, interacts with $N=4$ QH edge channels. The rate of tunnelling from one of the channels is studied as a function of the bias $\Delta\mu$ between the channel and the dot.}
\label{fig:system}
\end{figure}

Notably, the exact solution\cite{nozieres} demonstrates the universality of  FES exponents and the non-perturbative character of this effect. Such bright manifestations of interactions also occur in one dimensional (1D) Fermi systems, known as the Luttinger liquids.\cite{fradkin}
Among experimentally accessible configurations the quantum Hall (QH) effect  systems deserve a special attention. In this regime, the edge states of the two dimensional electron gases can be viewed as chiral 1D channels, whose direction of propagation is defined by the sign of the magnetic field. 
A convenient method to describe these states is to use the bosonization technique,\cite{giamarchi} that allows one to address the interactions non-perturbatively. Interestingly, this approach enables to find the FES power-law in a QH system, endowing it with a clear physical meaning. 
Namely, its manifestation was studied in tunnelling  to a single-level QD,\cite{jura} surrounded by a number of edge channels (see Fig.\ref{fig:system}). In the low-energy limit, the tunnelling rate dependence on the bias between the QD and the $m$th channel, where the electron tunnels from, acquires a form
\begin{equation}\label{powerlaw}
\Gamma \propto \Delta\mu^{\alpha}, \quad  \alpha = 2 q_m + \sum q^2_n,
\end{equation}
where $q_n < 0$ denotes the charge induced in the $n$th channel.
In this paper we would like to complete this picture, by considering the tunnelling rate at biases that reveal the structure of the energy levels of the QD. 

The bosonization approach provides means of describing the system in terms of scattering of bosons.\cite{sukhorukov} Being widely used\cite{cheianov, oreg2, artur, dario} for its clarity and relative simplicity, this formalism enables to look at the FES from another point of view by considering the dot as a compactified bosonic field. 
We apply this formalism in Sec.\ \ref{scatteringtheory} to derive the FES in the tunnelling rate in the low energy limit. We show that the scattering states, constituting the basis for the boson fields, get the meaning of positive charges in this limit. Surprisingly, those turn out to be the charges (with a negative sign), induced in the channels around the QD when it is charged, which leads to the universality of the FES phenomenon.

We then go beyond the single-level approximation for the QD in Sec.\ \ref{application}, so that the scattering of the incoming wave in the channel leads to an excitation of collective modes in the QD. 
The nature of these excitations is fully governed by two parameters: the coupling constant $\sigma = \sum_n q^2_n$ and the dispersion of plasmons in the QD. We first concentrate on the no dispersion case. Then, if there is no interaction, i.e. $\sigma = 0$, the tunnelling rate  behaves as a set of steps as a function of the bias. The steps correspond to the free fermion energy levels in the QD, implying that the bosonic and the fermionic pictures describe the same entity. If the interaction is ``turned on", the steps become smeared, due to the finite width that the energy levels acquire. The width is proportional to the coupling $\sigma$, but it also grows quadratically with the number of the energy level. Thus, even if the interaction is small, it leads to a non-perturbative effect, which we are able to describe analytically due to the correspondence between free-fermion levels and boson resonances in this geometry and the chirality of the boson fields.
When there is interaction in the QD, the spectrum of plasmons acquires in general a weak dispersion. We then demonstrate splitting of the fermion levels (starting from the third one). It originates from the shift between different single- and multi-plasmon processes corresponding to the excitation of a particular fermion level. The effect is mostly pronounced for the third level, on which we dwell in detail.

The following discussion returns to QDs with a linear spectrum for plasmons. In Sec.\ \ref{QDneutral}, we study the tunnelling rate to a QD with two chiral channels with strong long-range interaction. Those can be decoupled into charge and neutral modes. By a proper choice of the bias only the lowest energy level of the charged mode can be excited. However, the heights of the steps stop being equal and acquire a universal structure which is a consequence of the strong interaction between the channels, leading to the charge fractionalization. 

This result is immediately used to describe the rate of tunnelling to a three dimensional  QD in the Coulomb blockade regime in Sec.\ \ref{QDcoulomb}. Indeed, in such a case there is again a separation between the charge and neutral mode. Using the formalism from Sec.\ \ref{scatteringtheory}, we analyse this situation and show that the result is similar to that of the previous section. The difference is, however, in the fact that there is a direct coupling to the neutral modes so that the heights and the positions of the steps are arbitrary.

\section{Fermi - Edge Singularity from scattering theory}\label{scatteringtheory}

Let us consider a QD of the characteristic size $L$ interacting with $N$ edge channels of a QH system at integer filling factor. To see how the FES manifests itself, we use the bosonization technique \cite{giamarchi} and apply the scattering theory for bosons,\cite{sukhorukov} which simply accounts for interactions. The Coulomb interaction might be screened in a complicated way, therefore we do not assume any particular form of density-density interactions. We only require that the QD size $L$ is much smaller than the wave length of the density fluctuations in the edge channels (plasmons), thus taking into account the low-energy character of the FES. Note that we work in the units, where $\hbar = c = e = 1$. To bosonize the electrons in the QD, we consider its edge as a one-dimensional channel with the glued ends, i.e., forming a ring.
We express the electron operators in the channels, $\psi_n(x)$ and in the ring $\psi_0(x)$, with the help of the boson fields $\phi_n(x)$ and $\phi_0(x)$:
\begin{equation}
\psi_n(x) \propto e^{i\phi_n(x)},  n=0,...,N, 
\end{equation}
where the field $\phi_n(x)$ is related to the charge density operator $\rho_n(x) = \frac{1}{2\pi}\partial_x \phi_n(x)$. The commutator of bosonic fields $[\phi_n(x), \phi_m(y)] = i\pi\delta_{nm} \sgn (x-y) $ together with the above definition guarantees the fermion commutation relations and the electron charge $[\psi_n(x),\rho_m(y)] = \psi_n(x)\delta(x-y)\delta_{mn}$.

Next, we write down the Hamiltonian of the interacting fermions in terms of the new bosonic fields:
\begin{align}\label{hamiltonian}
\mathcal{H} = \mathcal{H}_0 + \mathcal{H}_{int} + \mathcal{H}_t,
\end{align}
where the free Hamiltonian is given by
\begin{equation}\label{freehamiltonian}
\mathcal{H}_0 = \frac{1}{4\pi}\sum_{n=0,..,N} v_n \int dx \left\lbrace \partial_x \phi_n(x)\right\rbrace  ^2.
\end{equation}
We included the interaction part
\begin{equation}\label{inthamiltonian}
\mathcal{H}_{int} = \frac{1}{8\pi^2} \sum_{nn'} \int \int dx dy U_{nn'}(x,y) \partial_x \phi_n(x) \partial_y \phi_{n'}(y)
\end{equation}
with arbitrary electrostatic potentials $U_{nn'}$. Finally, the last term describes tunnelling between the $m$th channel and the QD at some point $x_0 $:
\begin{align}\label{A}
\mathcal{H}_t = \tau(A + A^{\dagger}),\quad A =e^{i \left\lbrace \phi_0(x_0)-\phi_m(x_0)\right\rbrace}.
\end{align}
However, the exact position $x_0$ is of no interest because of the long wavelength limit, allowing to consider the bosonic field in a certain interaction region being independent of the coordinate. 

Next, let us calculate the tunnelling rate from one of the channels, say the $m$th one, to the QD when a bias $\Delta\mu$ is applied between them.\footnote{The rate of tunnelling in the opposite direction follows from the electron-hole symmetry.} Notably, the change of the electro-chemical potential at the channel shifts the dot level due to the electrostatic interaction. The value $\Delta\mu$ takes this effect into account and will be written down explicitly later.
Addressing the tunnelling term as a perturbation, one can express the tunnelling rate as the integral
\begin{equation}\label{current}
\Gamma = |\tau|^2 \int_{-\infty}^{\infty} dt\left\langle A(0)A^{\dagger}(t)\right\rangle.
\end{equation}
Note, that such an approach enables us to use the fact that the remaining part of the Hamiltonian, $H_0+H_{int}$, has a quadratic form in the bosonic fields. Then the tunnelling rate can be expressed in terms of their two-point correlators. The bosonic fields can be found from the equations of motion  $\partial_t{\phi_n(x)}=i \left[\mathcal{H}_0+\mathcal{H}_{int},\phi_n(x)\right], n = 0,...,N$ that reveal
\begin{align}\label{motion}
\partial_t \phi_n(t) &+ v_n \partial_x \phi_n(x) +\nonumber\\
 & +\frac{1}{2\pi} \sum_{n'=0,..,N} \int dy U_{nn'}(x,y)\partial_y \phi_{n'}(y) = 0.
\end{align}
The solution may be presented in the form
\begin{equation}
\phi_n( x, t)  = \varphi_n(x, t) + \delta\phi_n(x,t),
\end{equation}
where the zero mode $\varphi_n(x, t)$ and the fluctuating part  $\delta\phi_n(x,t)$ read 
\begin{align}
\varphi_n(x, t) & = -\mu_nt + \varphi^{(0)}_n(x),\label{zeromode}\\
\delta\phi_n(x,t) & = \int_0^{\infty} \frac{d\omega}{\sqrt{\omega}}\sum_{n'=1}^N\left( \Phi_{n'n\omega}(x) e^{-i\omega t}a_{n'}(\omega)+h.c.\right)\label{fluct}.
\end{align}
Basically, the zero modes solve the system \eqref{motion} in the zero frequency limit and satisfy the following equations:
\begin{align}\label{zeromodeeq}
\mu_n =  v_n \partial_x\varphi^{(0)}_n(x) + \frac{1}{2\pi} \sum_{n'}\int dy\partial_y\varphi^{(0)}_{n'}(y) U_{nn'}(x,y).
\end{align}
Thus, zero modes  describe stationary charge densities and corresponding phase shifts, while the deviations are taken into account by the fluctuating part. We expressed the latter in the second-quantized form in the basis of the scattering states $\Phi_{n'n\omega}(x)$ with the creation and annihilation operators $a^{\dagger}_{n'}(\omega)$, $a_{n'}(\omega)$ satisfying the usual bosonic commutation relations. The scattering states diagonalize the Hamiltonian $H_0+H_{int}$ and satisfy specific boundary conditions. Namely, the scattering state $\Phi_{n'n\omega}(x)$ is described by an incoming plane wave in the channel $n'$ which then scatters into all the other channels $n=1,\ldots,N$. Thus, the scattering state presents the set of $N+1$ functions, enumerated by the second index, while the first index enumerates the scattering states.

To find correlation functions entering the expression (\ref{current}) for the tunnelling rate, we exploit the low-energy limit and perform the perturbation expansion of the scattering states in vicinity of the QD in frequency:
\begin{equation}\label{exp}
\Phi_{nl\omega}(x) = \Phi^{(0)}_{nl}(x) + i \omega \Phi^{(1)}_{nl}(x).
\end{equation}
It means that we look at the asymptotic behaviour of the scattering states in the region $x \sim L$, for which our approximation $\omega L/v \ll 1$ is valid. On the other hand, such an expansion may be understood as a way to describe how strong the scattering is. Specifically, the  parameter $\omega L /v$ being small implies that almost the whole incident wave gets transmitted. Substituting now the expression \eqref{exp} into \eqref{fluct} and into \eqref{motion}, we arrive at the system
\begin{align}\label{expzero}
0 &= v_n \partial_x \Phi^{(0)}_{nl}(x) + \frac{1}{2\pi} \sum_{n'}\int dy  \partial_y \Phi^{(0)}_{nn'}(y) U_{ln'}(x,y),\\
\Phi^{(0)}_{nl}(x) &= v_n \partial_x \Phi^{(1)}_{nl}(x) + \frac{1}{2\pi} \sum_{n'}\int dy  \partial_y \Phi^{(1)}_{nn'}(y) U_{ln'}(x,y).\label{expfirst}
\end{align}
Obviously,  $\Phi^{(0)}_{nl}(x) = const$ are the solutions of \eqref{expzero}.
Particularly, for our scattering problem
\begin{align}\label{deltanl}
\Phi^{(0)}_{nl}(x) &= \delta_{nl}, \quad n,l =1,...,N ,\\
\Phi^{(0)}_{n0}(x)&=\varepsilon_0\label{eps0},
\end{align}
where $\varepsilon_0$ needs to be defined.

To clarify the physical meaning of $\Phi^{(0)}_{nm}$, we note that substituting \eqref{deltanl} and \eqref{eps0} into \eqref{expfirst} brings us to the same kind of electrostatic equations that identify the zero modes \eqref{zeromode}. So we may treat the problem of finding the coefficients $\Phi^{(0)}_{n0}(x)$ as an electrostatic one and formally write its solution
\begin{align}\label{formal}
q_n =\sum_{nn'} C_{nn'}\mu_{n'},
\end{align}
where $q_n = \frac{1}{2\pi}\int dx \partial_x\varphi_n^{(0)}(x),n = 0,1,...,N$ are the charges in the channels and at the dot. The particular form of the interaction is of no interest and is generally described by a capacitance matrix $C_{nn'}$. Taking into account \eqref{deltanl} and \eqref{eps0},  and assuming that all channels are grounded, except for the $n$th one, leads to:
\begin{equation}
q_n = C_{nn} + C_{n0} \varepsilon_0.
\end{equation}
As there is no ``charge'' at the dot $q_0 = 0$, it is easy to define its ``potential'' \footnote{It is worth noting that the potentials \unexpanded{$\mu_n$} in the electrostatic problem \unexpanded{\eqref{formal}} are of 1/time dimensionality, while the charges are dimensionless. However, in the scattering problem \unexpanded{\eqref{expzero}}-\unexpanded{\eqref{expfirst}} it is the potentials that are dimensionless and the charges get the time dimension. On the other hand, we are comparing the dimensionless potential \unexpanded{$\varepsilon_0$} and the charge \unexpanded{$q_n$}, so that the fact that they are equal is self-consistent.}
\begin{equation}\label{pot}
\varepsilon_0 = - C_{n0}/ C_{00}.
\end{equation}

On the other hand, if one poses a question on how the QD with a unit charge gets screened by  grounded channels, from $q_n = C_{n0}\mu_0 $ and $q_0=1$ one gets
\begin{equation}\label{static}
q_n =C_{n0}/C_{00}.
\end{equation}
We then conclude that the ``potential'' $\varepsilon_0$, induced in the dot in the particular situation where the plane wave is incident in the channel $n$ is the same up to a sign as the charge induced in this  channel in the set-up when the dot is charged:
\begin{equation}\label{connection}
\varepsilon_0 = -q_n, \quad q_n < 0.
\end{equation}
Thus, in the low energy limit the scattering and the electrostatic problems are simply connected, which reflects the universality of the FES phenomenon.
Finally, as it is mentioned above,  the interaction between the biased channel $m$ and the QD raises the dot's energy level by $\mu_0 =- \mu_m C_{m0}/C_{00}$ [compare to Eq. (\ref{pot})], so we denote the difference in their potentials as 
 $\Delta\mu = \mu_m-\mu_0$.

We now move to the calculation of the tunnelling rate \eqref{current} :
\begin{align}\label{gamma0}
\Gamma\propto \int_{-\infty}^{\infty} dt \exp\left\{-i \Delta\mu t +\mathcal{K}_1(t)+\mathcal{K}_2(t)\right\},
\end{align}
where we introduced the auto-correlator $\mathcal{K}_1(t)$ and the cross-correlator $\mathcal{K}_2(t)$ of the bosonic fields:
\begin{eqnarray}
 \mathcal {K}_1(t) &=& -\sum _{n={0,1}} \langle \left[\delta \phi_n (t) -  \delta \phi _n(0)\right]\delta \phi _n(t)\rangle ,
\\
\mathcal {K}_2(t)& =&\left \langle \delta \phi _n(0)\left[ \delta \phi _0(0)- \delta \phi _0(t)\right] \right \rangle
\nonumber\\&& \hspace*{1.5cm}- \left \langle \left[ \delta \phi _0(0) -\delta \phi _0(t)\right] \delta \phi _n(t) \right \rangle. 
\end{eqnarray}
To find them we use the spectral decomposition \eqref{fluct} of the fields in the channels and in the dot.
Thereby, we arrive at the following result for the correlators:

\begin{align*}
\mathcal{K}_1(t) &= -(1+\sum_{n=1}^{N} q^2_j) \int_0^{\infty} \frac{d\omega}{\omega} \left(1-e^{i\omega t}\right),\\
\mathcal{K}_2(t) &= -2q_m \int_0^{\infty} \frac{d\omega}{\omega} \left(1-e^{i\omega t}\right).
\end{align*}
Introducing a cut-off $\delta^{-1}$ at high energies, we calculate the above integral $\int_0^{\infty} \frac{d\omega}{\omega} \left(1-e^{i\omega t}\right)e^{-\delta \omega} = \log\left( \frac{\delta - it}{\delta}\right) $. Finally, the formula \eqref{gamma0} leads to the following outcome for the tunnelling rate from the $m$th channel to the dot: 
\begin{align}
\Gamma &\propto \int_{-\infty}^{\infty} dt \frac{e^{i\Delta\mu t}}{\left(\delta + it\right)^{1+ \alpha}} = \frac{2\pi \theta(\Delta\mu)}{\Gamma(1+\alpha)}\Delta\mu^{\alpha}\\
\alpha &= 2q_m + \sum_{n = 1}^{N} q^2_n, \quad q_m < 0.\label{alpha}
\end{align} 
Hence, we reproduced the FES power-law for a single level quantum dot in the long wavelength regime. At this point one might appreciate the simplicity of the used method, while in the next section we also demonstrate its applicability for a more complicated system.

To complete the formal description of our approach, we make some comments on the role of the interaction between the co-propagating channels before they get split close to the dot. The tunnelling rate is determined by the local correlators and, in fact,  it can be shown that the upstream interaction does not influence the local correlators in the vicinity of the QD. Namely, rewriting the boson fields in terms of the scattering states coming from the interacting region, one might see that the local correlators are the same as in the case with no interaction in the region upstream the QD at all.

\section{Application to the collective mode assisted tunneling}\label{application}

It has been demonstrated in the previous section that the FES physics is universal in the low energy limit. This is reflected in the particular power-law form of the tunnelling rate as a function of the bias where the exponent is only defined by the charges induced in the channels around the dot.
In this and following sections, we would like to study tunnelling to a multilevel QD  and discuss the manifestation of the FES in such a case. We continue to work within the same framework and set-up, but now assume that the characteristic size $L$ of the dot can be larger than the wave length of excitations $\lambda$, which in turn is much larger than the characteristic size $W$ of the interaction region, implying relatively large QDs: $L\gg W$. This allows us to consider tunnelling to excited states in a QD, nevertheless expecting the same level of universality as in the FES effect.

In this case, the local interaction of the QD with each channel can be described by introducing the scattering coefficients $r_n$ and $t_n$ for the reflection and the transmission of the plasmons, respectively. We then derive the scattering states $\Phi_{m0\omega}$ at the dot explicitly. The procedure is as follows. Recall that the bosonic fields in the channels in the proximity of the dot are approximated by \eqref{fluct}, \eqref{exp} and \eqref{deltanl} for $W \ll \lambda$, so the field in the $m$th channel has the following form:
\begin{align*}
\delta\phi_m(x,t) &= \int_0^{\infty}\frac{d\omega}{\sqrt{\omega}}\left(e^{-i\omega t}a_m(\omega)+h.c.\right) \\
&\equiv\int_0^{\infty}\frac{d\omega}{2\pi}\left(e^{-i\omega t}\delta\phi_m(\omega)+h.c.\right).
\end{align*}
The next step would be to express the field $\delta\phi_0(\omega)$ at the dot in the interaction region with the $m$th channel in terms of all the fields $\delta\phi_n(\omega), n=1,...,N$. There are two processes that define the value of $\delta\phi_0(\omega)$. Firstly, it is the reflection of the fields in $N$ channels to the dot and, secondly, the successive transmission of the field $\delta\phi_0(\omega)$  through all the interaction regions. 

\begin{figure}[h]
\includegraphics[scale=0.8]{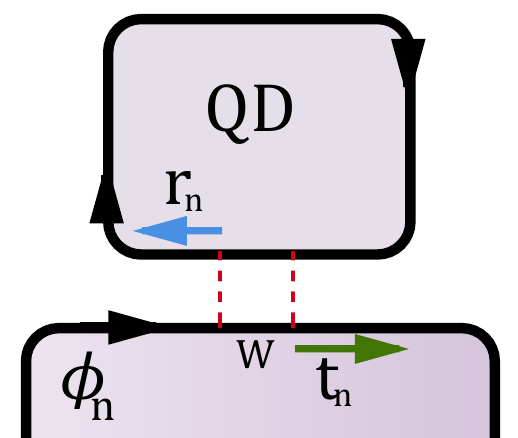} 
\caption{Sketch of the scattering process between the QD and the $n$th channel. The arrows with the corresponding scattering coefficients $r_n$ and $t_n$ schematically represent the direction of a boson mode propagation. }\label{scatteringpic}
\end{figure}

Note that the reflected field from the $n$th channel acquires a certain phase corresponding to the distance $L_{mn}$ between the contacts of the QD with this channel and the $m$th one, i.e., the channel from which the tunnelling occurs.
Making use of the scattering coefficients $r_n, t_n$ (see Fig.\ \ref{scatteringpic}), we  arrive at the following expression for the  field $\delta\phi_0(\omega)$:
\begin{align}
\delta\phi_0 = \sum_{n=1}^{N} r_n \delta\phi_n e^{i \frac{\omega}{v} L_{mn}}+ \prod_{n=1}^{N}t_n e^{i \frac{\omega}{v} L}\delta\phi_0.
\end{align}
We use this result to rewrite the field $\phi_0(\omega)$ in terms of the scattering states $\Phi_{0m\omega}$ :
\begin{align}\label{phidot}
\delta\phi_0(\omega) &= \sum_{n=1}^N \Phi_{0n\omega} \delta\phi_n(\omega),\\
\Phi_{0n\omega} &= \frac{r_n }{1-\prod_{n=1}^N t_i e^{i \frac{\omega}{v} L}}.\label{dotscatteringstate} 
\end{align}
Note, that we have ignored the phases $\exp(i\omega L_{mn}/v)$ as they cancel each other, which can be seen from the following expressions for 
the auto- and cross-correlators 
\begin{align}
&\mathcal{K}_1 = -\int_0^{\infty} \frac{d\omega}{\omega} \left(1-e^{i\omega t}\right)\left(1+\sum_{n=1}^N \vert\Phi_{0n\omega}\vert^2\right), \label{auto}\\
&\mathcal{K}_2 = 2\int_0^{\infty} \frac{d\omega}{\omega} \left(1-e^{i\omega t}\right)\Re(\Phi_{0m\omega}).\label{cross}
\end{align}

To analyse the form of the scattering states \eqref{dotscatteringstate} in the low energy limit, 
$\omega W/v\ll 1$, it is enough to expand the scattering coefficients up to the second order in frequency. Then using the unitarity of the scattering matrix we arrive at:
\begin{align}\label{unitarity1}
r_n &=  i\omega \tilde{r}_n - \omega^2 \tilde{r}_n \tilde{t}_n , \\
t_n &= 1 + i \omega \tilde{t}_n - \omega^2 \frac{\tilde{r}^2_n+\tilde{t}^2_n}{2}.\label{unitarity2}
\end{align}
To understand the physical meaning of the coefficients $\tilde{r}_n$ and $\tilde{t}_n$, we note that on one hand $\Phi_{0m\omega} (\omega = 0)= -q_m$ from \eqref{eps0} and \eqref{connection}, on the other hand $\Phi_{0m\omega} (\omega = 0) = \tilde{r}_m/\left(L/v +\sum_{n} \tilde{t}_n\right)$. Finally, recalling \eqref{static} it becomes evident that $\tilde{r}_m$ and $L/v +\sum_{n} \tilde{t}_n$ are just the capacitances (up to a constant multiplier) between the channels $C_{m0}$ and the self-capacitance $C_{00}$, respectively. We simply denote $\tau_C = L/v +\sum_{n} \tilde{t}_n$, since this is just the travel time of the plasmon in the dot.
With this in mind, and using Eqs.\ (\ref{unitarity1}) and (\ref{unitarity2}), we rewrite the expression for the scattering states in the form
\begin{equation}\label{phi}
\Phi_{0n\omega} = \frac{i\omega \tau_Cq_n}{1-(1-\frac{\sigma_0}{2}\omega^2 \tau_C^2)e^{i\omega \tau_C}},
\end{equation}
where we introduced the dimensionless coupling constant 
\begin{equation}\label{coupling}
\sigma = \sum_{n=1}^N q^2_n
\end{equation}
characterizing the strength of the interaction. 

 To find the auto-correlator \eqref{auto}, we start with analysing $|\Phi_{0n\omega} |^2$. Assuming that coupling is weak, $\sigma\ll 1$ (either due to the large number of channels, $N\gg  1$, or because of partial screening by a gate, $\sum_nq_n\ll 1$), the main contribution to the integral in \eqref{auto} comes from the singularity at $\omega=0$ and from the set of plasmon resonances in \eqref{phi} at frequencies $\omega_l=\Delta\omega l$, $l=1,2,\ldots$, where $\Delta\omega=2\pi/\tau_C$.
Evaluating these contributions separately, we can write
\begin{align}
&\mathcal{K}_1  = - \left(1 +\sigma\right)\log\left(\frac{\delta-it}{\delta}\right)-J,\label{autoJ}\\
&J = \sum_{l=1}^{\infty}\frac{e^{-\varepsilon l}}{l}\left[1 - \exp{\left(i l \Delta\omega t - \pi l^2 \sigma \Delta\omega |t|\right)}\right],\label{J}
\end{align} 
where $\varepsilon$ is the high-energy cutoff parameter.
Considering first the free-fermionic case of $\sigma=0$, the sum in \eqref{J} can be evaluated explicitly, and we obtain
\begin{equation}
J_{free} =  \log\left(1 - e^{i\Delta\omega t-\varepsilon}\right),
\end{equation}
where we dropped unimportant constant contribution. Therefore, the tunnelling rate \eqref{gamma0} is described by a set of steps as a function of a bias:
\begin{align}
\Gamma_{free} &\propto \int_{-\infty}^{\infty} dt\frac{e^{i\Delta\mu t}}{\left(\delta + i t\right)} \frac{1}{1 - e^{-i\Delta\omega t-\varepsilon}} \nonumber \\
&\propto\sum_{n=0}^{\infty}\theta(\Delta\mu-n\Delta\omega),
\label{Gammanoint}
\end{align}
where we set $\varepsilon=0$ in the end of calculations. 
This result is in perfect agreement with the free fermionic picture, as the steps correspond to the quantized energy levels of the QD. This happens because there is no interaction in the QD itself, as it is being screened. So the electronic and bosonic description are just the two alternative ways to look at the same system.

Returning now to the interaction case, instead of calculating the sum over $l$ in \eqref{J}, we formally represented $\exp(-J)$ as  a Taylor series and changed the sign of the integration variable $t$ for convenience:
\begin{align}
&\Gamma(\Delta\mu) \propto \int_{-\infty}^{\infty} dt \frac{e^{i\Delta\mu t+\mathcal{K}^{(1)}_2}}{\left(\delta + i t\right)^{1+\alpha}} \nonumber
\\
&\times \sum_{m=0}^{\infty}\frac{1}{m!} \left(\sum_{l=1}^{\infty}\frac{e^{-\varepsilon l}}{l} \exp(-i l\Delta\omega t -l^2\sigma \pi\Delta\omega |t|)\right)^m\label{Gammafinal2}.
\end{align}
Here, we also separated the Mahan term
\begin{equation}\label{mahan}
\mathcal{K}^{(0)}_2 = -2q_m \int_0^{\infty}\frac{d\omega}{\omega} \left(1-e^{-i\omega t}\right) = -2 q_m \log\left(\frac{\delta+it}{\delta}\right)
\end{equation}
from the cross-correlator 
$\mathcal{K}_2(-t)=\mathcal{K}^{(0)}_2+\mathcal{K}^{(1)}_2$
to complete the FES exponent $\alpha$ in the denominator.
We show  in the appendix \ref{appendix2} that the term $\mathcal{K}^{(1)}_2$ is negligible, as it can be considered perturbatively in the coupling $\sigma$.

\begin{figure}[h]
\includegraphics[scale=0.4]{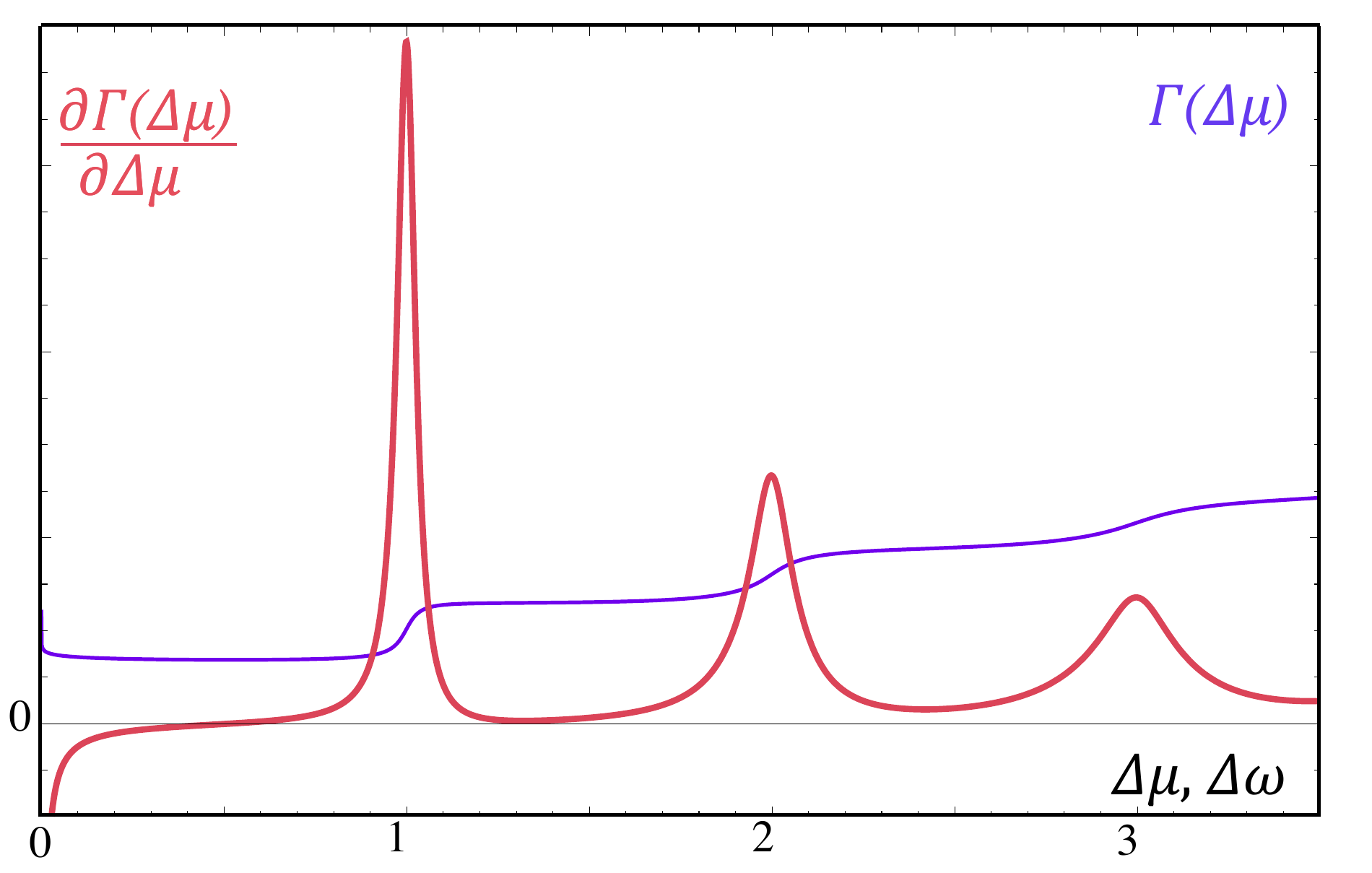} 
\caption{The tunnelling rate $\Gamma(\Delta\mu)$ and its derivative $\partial\Gamma(\Delta\mu)/\partial\Delta\mu$  is shown for the FES exponent $\alpha=-0.04$  and the coupling constant $\sigma = 0.01$. It is evaluated approximately according to \eqref{Gammafinal2} by keeping the finite number of terms. The values are presented in arbitrary units, as we study them up to  constant multipliers.}
\label{fig:chargemode}
\end{figure}

After opening the brackets in Eq.\ \eqref{Gammafinal2}, each term in the triple sum is of the form $const \cdot \exp(-i n \Delta\omega t - p \sigma \Delta\omega |t|)$, $n$ and $p$ being integers, and thus contributes to the $n$th step in the function $\Gamma(\Delta\mu)$. Recalling that $\sigma\ll 1$, and that we are interested in low energies, we can simply present $\Gamma(\Delta\mu)$ as a sum over steps\footnote{Formally, the time integral in Eq.\ (\ref{Gammafinal2}) does not vanish for $\Delta\mu<0$, because the integrand is not analytical function. Moreover, the sum inside the integral seems to be divergent. This problem is related to our approximate way of evaluating the contributions of the plasmon resonances in Eq.\ (\ref{auto}), which correctly describes the steps in $\Gamma(\Delta\mu)$ but fails far away from the steps.  One can check that when expanding $\exp(-J)$ the phase space limitation in multiple integrals over $\omega$ solves this problem, and that correction to our result is small in $\sigma$. }
\begin{equation}\label{gammaanalytic}
\Gamma(\Delta\mu) = \frac{2\pi}{\Gamma(1+\alpha)}(\Delta \mu)^{\alpha} +\sum_{n=1}^{n_0}\Gamma_n,\quad \Delta\mu>0,
\end{equation}
and cut the sum at arbitrary $n_0>\Delta\mu/\Delta\omega$. Here the first term is the FES contribution. Proceeding in this way, we  evaluate the integral \eqref{Gammafinal2} term by term and present  the result in terms of the functions 
\begin{equation}
G(\gamma, \Delta\mu_n ) = \frac{2\pi\Im\left\{(\gamma\Delta\omega+ i \Delta\mu_n )^{\alpha}e^{i\pi\alpha/2}\right\}}{\sin(\pi\alpha)\Gamma(1+\alpha)},\label{G2}
\end{equation}
where we define $\Delta\mu_n\equiv\Delta\mu-n\Delta\omega $. Thus, e.g., the first three steps can be presented as
\begin{align}
&\Gamma_1 = G(\pi\sigma, \Delta\mu_1),\nonumber\\
&\Gamma_2 =  \frac 12 G(2\pi\sigma, \Delta\mu_2)+\frac 12 G(4\pi\sigma, \Delta\mu_2),\label{gammas}\\
&\Gamma_3 = \frac 13 G(9\pi\sigma, \Delta\mu_3) + \frac 12 G(5\pi\sigma, \Delta\mu_3) + \frac 16 G(3\pi\sigma, \Delta\mu_3).
\nonumber
\end{align} 
The result of such evaluation is shown in Fig.\ \ref{fig:chargemode}.

Quite remarkably, the results (\ref{gammaanalytic}-\ref{gammas}) show that single-electron levels in the QD can be viewed as  plasmon resonances. Indeed, although each term in Eqs.\ \eqref{gammas} for $\Gamma_n$ represents single- or multi-plasmon process, and their broadening is caused by Coulomb interaction, in the free-fermionic limit $\alpha,\sigma\to 0$ they add to the single-electron excitation step: $\Gamma_n=\theta(\Delta\mu-n\Delta\omega)$. This is a consequence of the fact that we assume the linear spectrum of the plasmon in the QD, which is known to leave electrons effectively free. This leads us to the next idea to relax this limitation by considering a weak dispersion in the spectrum of plasmons. The immediate consequence of this is that one should expect splitting of the steps in $\Gamma$ starting from $n=2$, where it is also most pronounced. 

\begin{figure}[h]
\center
\includegraphics[scale=0.4]{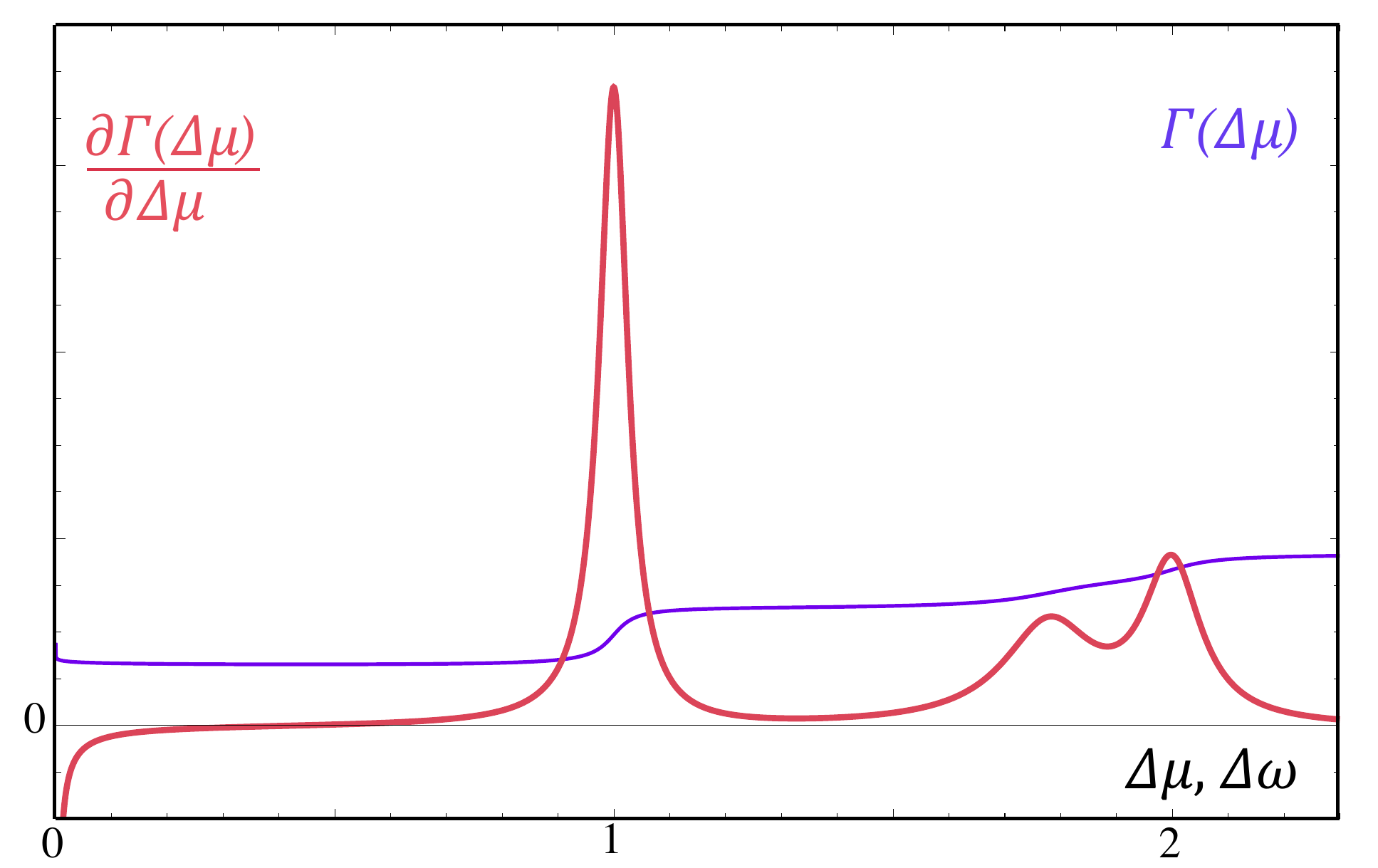} 
\caption{Splitting of the second electron level in two levels corresponding to one- and two-plasmon processes in the case of weakly dispersive plasmons is shown.
The tunnelling rate $\Gamma(\Delta\mu)$ and its derivative $\partial\Gamma/\partial(\Delta\mu)$ are calculated using Eqs.\ \eqref{G2}-\eqref{gammasplit3} for $\alpha = -0.02$, $\sigma = 0.01$ and $\epsilon = 0.22$.}
\label{fig:splitting}
\end{figure}

Indeed, considering the situation where the Coulomb interaction in the QD is not screened by a gate, it is natural to assume the spectrum of plasmons to be weakly concave,\cite{PhysRevB.85.075309} so that the frequency of the second plasmon resonance
acquires a small negative shift.
To describe the first two peaks as well as the ground state analytically, we make use of an expression \eqref{gammaanalytic}, introducing a small shift: $2\Delta\omega\to 2\Delta\omega-\epsilon $. Since two contributions to $\Gamma_2$ in Eq.\ \eqref{gammas} represent single- and two-plasmon processes, this splits the second step in two steps:
\begin{equation}
\Gamma_2 = \frac 12 G(2\pi\sigma, \Delta\mu_2) + \frac 12 G[(2-\epsilon)^2\pi\sigma, \Delta\mu_2 + \epsilon]\label{gammasplit3},
\end{equation}
while $\Gamma_1$ and the FES contributions remain unchanged. 
Here, the lower energy step  corresponds to the emission of one plasmon with the energy $2\Delta\omega-\epsilon$, while the higher energy step refers to the emission of two plasmon of  the energy $\Delta\omega$.  The width of the former one is larger and is defined by $(2-\epsilon)^2\pi \sigma \Delta\omega$ as compared to $2\pi \sigma \Delta\omega$ for the higher peak. The results of the calculation are shown in Fig.\ \ref{fig:splitting}. Higher peaks, not shown in this figure, will show additional splittings corresponding to the number of emitted plasmons. If the coupling
constant is small enough, $\sigma<\epsilon/\Delta\omega$, the effect should be clearly seen in experiment.

\section{Tunnelling to a QD with neutral modes}\label{QDneutral}

One of the conclusions that one can draw from the results of the previous section is that in the case of the linear plasmon spectrum the effects of interaction are controlled roughly by a single parameter, coupling constant $\sigma$. One one hand, this leads to the universality of the effect of level broadening, which is a non-perturbative interaction effect reflecting a subtle electron-plasmon correspondence. On the other hand, 
smearing of the steps in $\Gamma(\Delta\mu)$ results also in the suppression of the FES effect in the excited states, so that it is observable only in the transition to the ground state. Below, we demonstrate that including neutral modes in the QD helps to circumvent this limitation. 

Let us consider the QD with two chiral channels with strong long-range interaction. \citep{2ch} Without loss of generality of our main conclusion below, we can concentrate on the case of only one external channel. We denote the bosonic fields as $\phi_1(x)$ for the channel in the lead and $\phi_2(x)$, $\phi_3(x)$ for the outer and inner channel at the QD, respectively. Again, we study the tunnelling rate to the QD according to \eqref{current}, specifically, to its external channel. The interaction between the two channels at the QD can be``turned off'' by a standard rotation of the basis: 
\begin{equation}\label{rotation}
\phi_2 = \frac{1}{\sqrt{2}} \left( \tilde{\phi}_2 + \tilde{\phi}_3\right),
\end{equation}
where $\tilde{\phi}_2(x)$ is the fast charge mode, and $\tilde{\phi}_3(x)$ is the slow dipole (neutral) mode. Hence, the tunnelling operator in \eqref{A} will include both charge and neutral fields. 

In the case of a strong interaction that we are dealing with, the neutral modes are much slower than the charge ones. That allows to consider biases $\Delta \mu$  smaller than the level spacing of the charge mode, so that only the neutral mode is excited.
Formally, this can be expressed similarly to Eq.\ \eqref{dotscatteringstate}:
\begin{align}
\delta\tilde{\phi}_2(\omega) = \frac{r_1}{1-t_1} \delta\phi_1(\omega), \quad
\delta\tilde{\phi}_3(\omega) = \frac{r_2}{1-t_2 e^{i k L}}\delta\phi_1(\omega),
\end{align}
where $r_{1,2}$ and $t_{1,2}$ are the scattering coefficients. They can be expanded as in Eqs.\ \eqref{unitarity1} and \eqref{unitarity2}, which guarantees  the unitarity fo the scattering matrix.  Using the analog of Eqs.\ \eqref{eps0} and \eqref{connection}, one can write $\delta\tilde{\phi}_2 = -\sqrt{2} q\delta\phi_1$,
\footnote{The additional factor of $\sqrt{2}$ in the case of two edge channels appears as a result of renormalization of the potential \unexpanded{$\eps_0$} in  Eq.\ (\ref{formal}) corresponding to one edge channel in the QD. However, it cancels with $\sqrt{2}$ from the unitary transform (\ref{rotation}). It leads, in the end, to the proper form of the FES exponent, as it should, due to the universality of this phenomenon.} where the charge $q<0$ is induced in the channel outside the dot. 

Repeating now the steps that lead to Eqs.\ \eqref{autoJ} an \eqref{J}, we observe that the charge mode contributes to the low-energy part of the correlator in \eqref{autoJ} with the coupling constant $\sigma_c=q^2$, while the neutral mode mostly contributes to the term \eqref{J} with its own coupling $\sigma_n$ to the field $\varphi_1$, which is, by all means, significantly smaller than for the charge mode. That is to say there is almost no interaction between them. Therefore, in first approximation, one can set $\sigma_n=0$, and the sum in \eqref{J} can be evaluated explicitly (as in the case of free electrons) with the result
\begin{align}
\Gamma(\Delta\mu) &\propto \int_{-\infty}^{\infty} dt\frac{e^{i \Delta\mu t}}{(\delta + i t)^{1 +\alpha}}\frac{1}{(1-e^{i\Delta\omega t-\varepsilon})^{1/2}}\nonumber \\ 
&\propto 
\sum_{n=0}^{\infty}c_n \theta\left(\Delta\mu-n\Delta\omega\right)(\Delta\mu-n\Delta\omega)^{\alpha}\label{Gammaneutral},
\end{align}
where the FES exponent $\alpha=2q+q^2$, according to its universality, and $c_n=(-1)^nC_{-1/2}^n$. 

\begin{figure}
\includegraphics[scale=0.4]{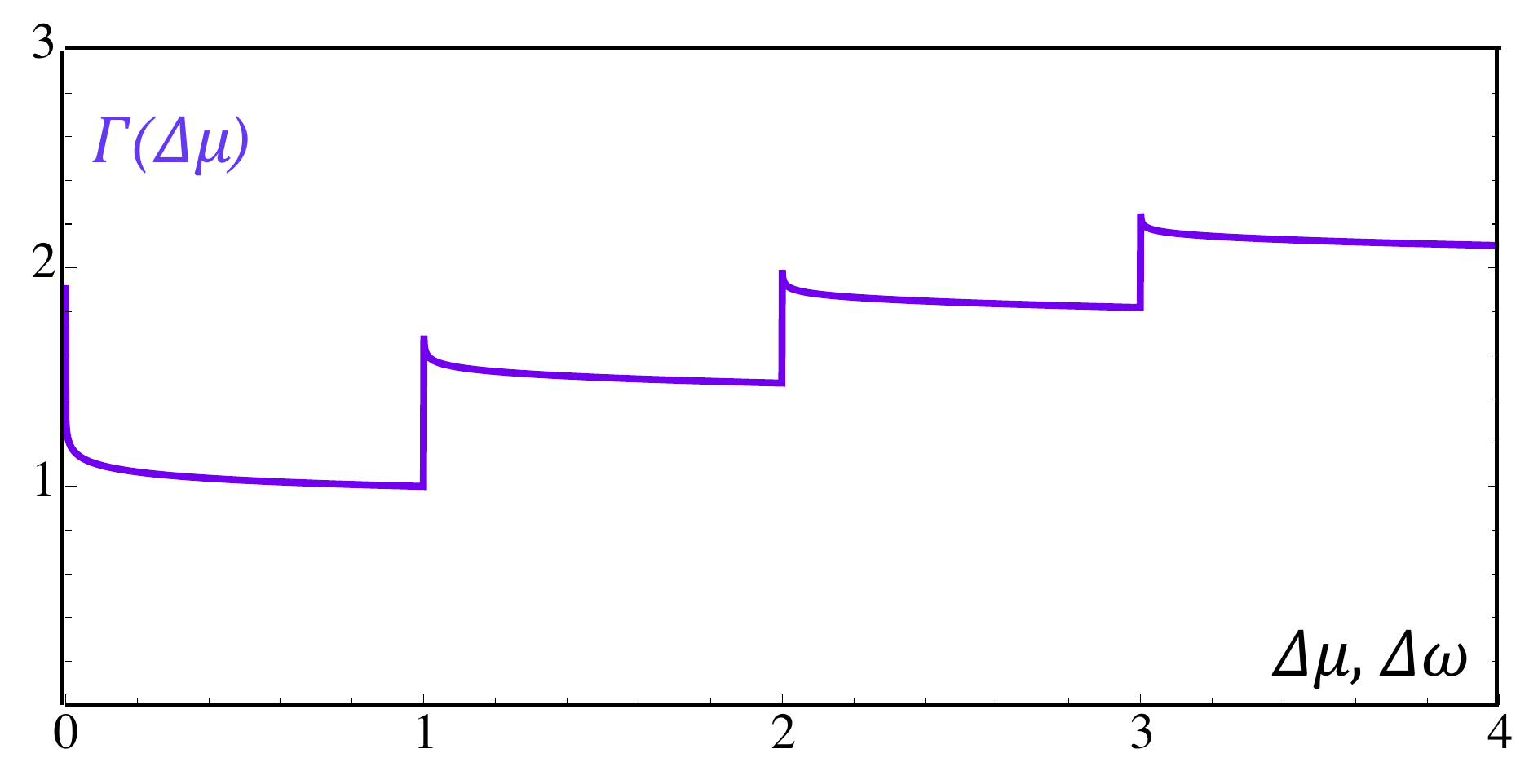} 
\caption{The tunnelling rate $\Gamma(\Delta\mu)$(in arbitrary units) calculated according to Eq.\ \eqref{Gammaneutral} with the coupling to the charge mode $\alpha = -0.04$. It represents the case of tunnelling from a QH channel to a QD with two  chiral channels strongly interacting via a long-range potential. In this case the Hamiltonian can be diagonalized in the bias of a charge and a neutral modes. The bias can be chosen to be smaller than the energy of the first excited state for the charge mode. Then, the steps are explained by the excitation of the resonances of the neutral mode, while the interaction with the charge mode guarantees the appearance of the FES at each edge of the steps. }
\label{fig:neutralpic}
\end{figure}

The obtained result is strikingly different from the case of a strong interaction with a charged mode alone. Now the $\Gamma(\Delta\mu)$ curve gains a structure of steps that correspond to the excitation of the neutral mode. But in addition, each step carries a power-law singularity at its edge, i.e., the FES replicates itself as can be seen in the Fig.\ \ref{fig:neutralpic}. This happens because the neutral excitations are almost completely separated from the charge and there is nothing to smear the effect besides the temperature. So the energy is mostly spent on exciting the dipole mode, while a small excess part of it gives rise to long-range electron-hole excitations, which are responsible for the FES phenomenon. 

We would like to mention also that in contrast to the fee-electron case,  the amplitude of the steps is weighted with the universal numbers $c_n$, which are independent of the details of the interaction. For example, first four numbers are equal to $1$, $1/2$, $3/8$, and $5/16$. These numbers originate from the expansion of the function  $(1-e^{i\Delta\omega t-\varepsilon})^{-1/2}$  in \eqref{Gammaneutral} as a Taylor series, while the square root can be viewed as a manifestation of the charge fractionalization caused by the strong interaction between channels in the QD.

\section{Tunnelling to a QD in the Coulomb blockade regime}\label{QDcoulomb}

In the previous section we have considered the situation, where the charge mode in the QD is separated by a Coulomb gap from the neutral mode, which is only weakly coupled to the charge mode outside the dot. However, because of the reduced dimensionality of the system and as a result of strong interactions, an electron is equally coupled to both modes, which has certain important consequencies, as discussed above. Here we consider another quite common situation, where the QD is formed by a metallic or a semiconductor granula of small size, so that the charged mode is again separated by the Coulomb gap which results in the Coulomb blockade effect. However, this time the QD is a 3D system with the consequence that an electron has  direct coupling to neutral modes. In order to analize this situation, below we use the formalism intruduced in Sec.\ \ref{scatteringtheory} to derive the analog of the well-known $P(E)$-theory.\cite{nazarov} 

Having said that we now consider the same QH system with a QD, described by the following Hamiltonian :
\begin{align*}
\mathcal{H} = \mathcal{H}_0 +\mathcal{H}_{int}  + \mathcal{H}_d + \mathcal{H}_t,
\end{align*}
with $ \mathcal{H}_0$ corresponding to the free Hamiltonian \eqref{freehamiltonian} of the QH channels. To model the 3D character of the dot and the Coulomb blockade, we compactify one of the bosonic fields $\phi_n(x)$ in \eqref{inthamiltonian} and take the zero limit for its length, $L\to 0$, which brings us to the following form of the interaction Hamiltonian
\begin{align}
\mathcal{H}_{int} = &\frac{1}{8\pi^2} \sum_{nn'} \int \int dx dy U_{nn'}(x,y) \partial_x \phi_n(x) \partial_y \phi_{n'}(y) \nonumber\\
&+ q\sum_{n} \int \frac{dx}{2\pi} U_{n}(x) \partial_x\phi_n(x)
\end{align}
and the QD's energy
\begin{equation}
\mathcal{H}_d  = \sum_k \varepsilon_k d^{\dagger}_k d_k + \frac{q^2}{2C},
\end{equation}
where $q$ is a charge at the QD 
and the second term corresponds to the charging energy.
Finally, the tunnelling Hamiltonian has the form 
\begin{equation}\label{tunQD}
\mathcal{H}_t = A + A^{\dagger}, \quad A = \sum_k \tau_kd^{\dagger}_k e^{i(\phi_0 - \phi_m)},
\end{equation}
so that the tunnelling occurs from the $m$th channel to the $k$th level of the QD. 

The field $\phi_0(x, t)$ represents the charged mode at the dot. However, as we consider the Coulomb blockade regime, i.e., the limit $L\to 0$, only a zero-mode \eqref{zeromode} as well the lowest energy mode in $\delta\phi_0(\omega)$ can be excited. The latter corresponds to the excitation of the ``zero'' scattering states $\Phi^{(0)}_{0n}$ defined in \eqref{eps0} and \eqref{connection}. 
Therefore, to calculate the tunnelling rate we, basically, need to repeat the calculations in Sec.\ \ref{scatteringtheory}. Thus, the part of the tunnelling Hamiltonian \eqref{tunQD} responsible for the charge mode reveals the FES contribution, while the neutral modes correlators lead to the appearance of the steps:
\begin{equation}
\Gamma(\Delta\mu) \propto
\sum_k |\tau_k|^2\theta\left(\Delta\mu-\varepsilon_k\right)(\Delta\mu-\varepsilon_k)^{\alpha}.
\end{equation}
As in the case of a QD with neutral modes considered in the previous section, we see FES effect replicated at each step corresponding to the excitation of a neutral mode. However, the important difference is that now neither the level spacing, nor the amplitudes of the steps are regular and universal functions.

\section{Conclusion}

Interaction of the QD with the QH edge channels results in various curious phenomena, that manifest themselves in the form of the tunnelling rate to the dot. Working in the framework of boson scattering theory we managed to describe them modelling the QD by a compactified boson field. We first demonstrated a particular convenience of this approach in describing  a well-known FES phenomenon at low energies. Its universality can now be understood as a consequence of the connection between the scattering problem in the low energy limit and the electrostatic problem of screening. Namely, the charges induced in the channels around the charged dot due to the interaction define the scattering states. 

This method also allowed us to go to higher energies and consider the excitation of the collective modes in the QD, which is fully controlled by the interaction with the external channels and in the QD itself. If the interaction inside the dot is screened, so that the spectrum of plasmons is linear, there is a full correspondence between the free-fermion levels and the plasmon resonances. Thus, in the case of no interaction between the channels and the QD, the tunnelling rate curve versus the bias is simply described by the set of steps. Whereas in the presence of interaction, the steps become smeared. It is important to note, that though we consider a relatively small coupling, the effect is non-perturbative. However, it was possible to describe the analytical solution due to the fermion-boson correspondence as well as the chirality of bosons.
Next, we elaborated on the case of a weak dispersion in the spectrum of plasmons in the QD which leads to the splitting of fermion levels. This rather complicated behaviour can be again easily explained in terms of plasmons. 

Finally, a QD with two chiral strongly interacting edge channels also reveals interesting physics. Describing the system in terms of well separated charge and neutral modes we showed that the tunnelling rate acquires a universal form  at low enough energies. Remarkably, a different set-up with a QD in Coulomb blockade regime, exhibits a similar behaviour  and can be treated within our general approach. Nevertheless, unlike the previous case, a direct coupling to the neutral mode leads to the non-universal structure of the result. All the discussed phenomena can be explored experimentally, which we strongly recommend.

\section*{Acknowledgements}

We thank Ivan Levkivskyi for useful discussions. This work was supported by the Swiss National Science Foundation.

\appendix*

\section{Crosscorrelator for the multilevel QD}\label{appendix2}

Let us justify why $\mathcal{K}^{(1)}_2$ can be considered  as a perturbative correction in Eq.\ \eqref{Gammafinal2}. We start with applying the expression for the scattering states \eqref{phi} to \eqref{cross} to obtain the cross-correlator
\begin{align}
&\mathcal{K}_2 = -2q_m\int_0^{\infty}dx\left(1-e^{ixt/\tau_C}\right)\frac{\sin x}{|\Delta|^2}\label{crossapp},\\
&\Delta = 1-[1-(i\sigma/2) x^2]e^{ix},
\end{align}
where we introduced the dimensionless variable $x=\omega\tau_C$.
The integral \eqref{crossapp} has a logarithmic divergence for small $x$ cut by $e^{-i x t /\tau_C}$. We explicitly singled out this divergence in  Eq.\ \eqref{Gammafinal2} to find the tunnelling rate. This is the Mahan contribution that defines the correct power-law exponent of the FES. The rest of the cross-correlator can be written as
\begin{equation}\label{crosscorrection}
\mathcal{K}^{(1)}_2 = -2q_m\int_0^{\infty}dx\left(1-e^{-ix\frac{t}{C}}\right)\left(\frac{\sin x}{|\Delta|^2} -\frac{1}{x}\right).
\end{equation}
This part is proportional to $q_m \ll 1$ and here we demonstrate that the integral itself is of the order one.

 First and foremost, we make sure that there are no more divergences.
We eliminated the divergence for small $x$, but there is an ultra-violet cut-off at large $x$. However, this contribution is also  cancelled by the same cut-off in the FES [see Eq.\ \eqref{mahan}].
We also note that the term containing the time exponent vanishes for large values of $x$ due to strong oscillations. 
Next, we have to check that by isolating $\mathcal{K}^{(0)}_2$, the Mahan term, we did not lose any additional contributions coming from the small $x$. Indeed, expanding around zero $\frac{\sin x}{|\Delta|^2}-\frac{1}{x} \sim\frac{1}{x-x^3 \sigma/2} -\frac{1}{x} \sim  x \sigma/2 $, one can see that the deviation is small in $\sigma$.
Finally, concerning the constant part of the correction, it can be left aside as we are describing the tunnelling rate up to a multiplier.
Consequently, the term $\mathcal{K}^{(1)}_2 $ in the cross correlator represents a perturbative correction in coupling in the next order and thus can be neglected.


\bibliographystyle{apsrev4-1}
\bibliography{fermiedge}

\end{document}